\documentclass[sigconf,authorversion,nonacm]{acmart}
\AtBeginDocument{%
  }

\usepackage{graphicx}
\begin{document}
\begin{sloppy}

\title{Towards the First Code Contribution:\\Processes and Information Needs}

\author{Christoph Treude}
\affiliation{
  \institution{Singapore Management University}
  \city{Singapore}
  \country{Singapore}
}
\email{ctreude@smu.edu.sg}

\author{Marco A. Gerosa}
\affiliation{
  \institution{Northern Arizona University}
  \city{Flagstaff, AZ}
  \country{United States}
}
\email{marco.gerosa@nau.edu}

\author{Igor Steinmacher}
\affiliation{
  \institution{Northern Arizona University}
  \city{Flagstaff, AZ}
  \country{United States}
}
\email{igor.steinmacher@nau.edu}

\begin{abstract}
Newcomers to a software project must overcome many barriers before they can successfully place their first code contribution, and they often struggle to find information that is relevant to them. In this work, we argue that much of the information needed by newcomers already exists, albeit scattered among many different sources, and that many barriers can be addressed by automatically identifying, extracting, generating, summarizing, and presenting documentation that is specifically aimed and customized for newcomers. To gain a detailed understanding of the processes followed by newcomers and their information needs before making their first code contribution, we conducted an empirical study. Based on a survey with about 100 practitioners, grounded theory analysis, and validation interviews, we contribute a 16-step model for the processes followed by newcomers to a software project and we identify relevant information, along with individual and project characteristics that influence the relevancy of information types and sources. Our findings form an essential step towards automated tool support that provides relevant information to project newcomers in each step of their contribution processes.
\end{abstract}



\maketitle

\section{Introduction and Motivation}

When developers join a software development project, they find themselves in an unfamiliar project landscape~\cite{Dagenais2010} and have to overcome many barriers, including unclear or incomplete documentation, not being able to find their way to start, and technical hurdles~\cite{fronchetti2023contributing, rehman2020newcomer, Steinmacher2014}. These barriers exist in open source~\cite{Steinmacher2015} as well as in industry projects~\cite{Begel2008, Rastogi2015}.

Arguably, many of the barriers encountered by newcomers to software projects can be addressed by producing documentation that is specifically aimed at and customized for newcomers. Although the increase in social media use by software developers~\cite{fang2020need, Storey2010} has led to a plethora of documentation available online for virtually any software product~\cite{Treude2012}, this documentation is often poorly structured, incomplete, and scattered throughout different places and formats~\cite{fronchetti2023contributing}. In the words of one of our participants in this study: \textit{``Using Google is not always the easiest... there's a lot of people talking about it on all of these websites but actually trying to find the code or the documentation is buried or it redirects to pages that just no longer exist. [...] That's probably the main thing that I struggle with a lot''}.

To make some of this vast amount of documentation available in a format amenable to newcomers in open source projects, Steinmacher et al.~developed FLOSScoach~\cite{Steinmacher2016,steinmacher2018let}, a web portal that presents documentation aimed at newcomers for a small set of open source projects. This documentation was manually selected and added to the portal. Our ultimate goal is to automate the creation of such a portal, i.e., to automatically identify, extract, generate, summarize, and present documentation that is relevant to newcomers, inspired by previous work on the successful repackaging of software artifacts, including bug reports~\cite{Czarnecki2012, Mani2012, liu2020bugsum, Rastkar2014,koh2021bug} and source code~\cite{Haiduc2010, McBurney2014, Moreno2013, haque2022semantic,zhang2023ga, geng2023interpretation}.

In order not to overwhelm newcomers and only present them with relevant information, we need a detailed understanding of the processes they follow and their information needs, as well as the project and developer characteristics that influence the particular processes and information needs. We define a \textit{newcomer} as a developer who is new to a particular project, as opposed to a \textit{novice} who would be new to software development in general.\footnote{While many of our participants in this study are experienced developers, many of them had been newcomers to specific projects recently.} We pose the following three research questions:

\begin{description}
\item[RQ1] What processes do newcomers follow to make their first contribution after entering a project?
\item[RQ2] What information is important to newcomers during this period?
\item[RQ3] How do project and individual characteristics relate to the processes followed by newcomers and the types and sources of information consulted?
\end{description}

To answer these questions, we conducted a survey with about 100 practitioners from open source and industry, asking them about their own experience as newcomers in their respective projects and their recommendations for other newcomers. Based on the insights provided by our participants, in this paper, we contribute a model for the processes followed by newcomers to a software project and we identify information types and sources relevant to newcomers. The process is made up of 16 steps with 2 cross-cutting strategies that we validated in follow-up interviews. Furthermore, we find that information on tools, technologies, and programming languages is significantly more important in projects with few newcomers and that the relevancy of specific information sources depends on the characteristics of the individual developer and the project.

Many researchers have investigated what motivates developers to work on a project (see~\cite{Crowston2008} for an overview), and much is known about the socialization process and motivations that developers follow from their first contribution onward (e.g.,~\cite{Jergensen2011,gerosa2021shifting}). To fill the gap between these two phases, we explicitly focus on supporting newcomers from after they have been assigned to a project or decided to contribute to it until they have made their first code contribution. Unlike Steinmacher et al.~\cite{Steinmacher2015} who focused on what hinders newcomers during this time, we take a more constructive approach focusing on the documentation newcomers need, from which sources it is available, and which steps need to be supported. This information is essential in formulating requirements for automated tool support to help newcomers place their first code contribution and for communities to organize their documentation and be better prepared for receiving newcomers.

The remainder of this paper is organized as follows: Section 2 reviews related work to frame our study within the existing literature. Section 3 describes our research methodology, including study design and data analysis techniques. Section 4 presents the results, highlighting key findings regarding newcomer information needs in software projects. Section 5 discusses these findings, considering their implications for practice and research. Section 6 outlines the limitations of the study, ensuring a transparent reflection on its scope. Finally, Section 7 concludes with a summary of our contributions and suggestions for future research directions.

\section{Related Work}

Our research contributes to the large body of work on the information needs of software developers, as well as the work on the processes followed by developers when encountering a new project and recommender systems that have been developed for them.

\subsection{Developers' Information Needs}

There has been a substantial amount of work on the information needs of software developers after the newcomer stage. For example, Fritz and Murphy~\cite{Fritz2010} provided a list of questions that focus on issues that occur within a project. Sillito et al.~\cite{Sillito2006} provided a similar list focusing on questions during evolution tasks. LaToza and Myers~\cite{LaToza2010} found that the most difficult questions from a developer's perspective dealt with intent and rationale. In their study on information needs in software development, Ko et al.~\cite{Ko2007} found that the most frequently sought information included awareness of artifacts and colleagues. Previous work on the Question and Answer website Stack Overflow~\cite{Treude2011b} found ``how-to'' questions to be the most common type of questions. A few authors focused on recommending previously answered questions or finding duplicates as a way to support developers~\cite{lill2024helpfulness, koswatte2021optimized, kamienski2023analyzing, lima2023looking}. Still, several authors have applied information foraging theory to software engineering~\cite{Burnett2014, Fleming2013, kuttal2021end, Niu2013, leon2023comparing, bradley2022sources}, a theory that assumes that, given a plethora of irrelevant information, humans have evolved strategies to efficiently find relevant information for their needs without processing everything, i.e., reducing the mental cost to achieve their goals~\cite{Lawrance2013}.

Such theories are not concrete enough to determine requirements for tool support in terms of the steps newcomers go through, what information they need on their way toward their first code contribution, and which sources this information is available from.

\subsection{Newcomers' Processes}

Regarding the processes followed by a newcomer to a software project, the work most closely related to ours is that by von Krogh et al.~\cite{Krogh2003}. They proposed a ``joining script'' for developers who want to participate in a project, which includes activities such as ``Express interest to contribute'' and ``Self-introduction''. Although there is some overlap with the steps that emerged from our analysis, most of the steps in the joining scripts are not actionable due to steps such as ``Repeated interest to contribute''. As a result, the steps identified in our work can be more easily supported by tools and documentation and are less dependent on community and social factors. Furthermore, while Steinmacher et al.~\cite{Steinmacher2014} provide a list of barriers such as ``lack of commitment'', these are arguably difficult to overcome with tool support alone; and the types of information, sources and processes proposed in FLOSScoach~\cite{steinmacher2018let, Steinmacher2016} are not grounded in empirical data. Han et al.~\cite{han2024understanding} focused on the onboarding to deep learning projects, and mapped the activities performed by newcomers based on the traces in the software repositories. Although this maps the steps of newcomers in a project, the contribution-centric approach does not cover the steps and activities that are not traceable, which are important to understand how the onboarding happens. Gregory et al.~\cite{gregory2020onboarding} studied the onboarding of newcomers to agile teams, and identified the practices adopted and challenges faced by newcomers when onboarding. Another study has investigated the role of onboarding programs in helping software developers place their first contribution (e.g.,~\cite{Labuschagne2015}); these programs can be seen as complementary to the goal of our work.

Herraiz et al.~\cite{Herraiz2006} tracked the activities of newcomers and found two groups with clearly different joining patterns: volunteers tend to follow a step-by-step joining process starting with an email message and a bug report followed by a bug fix and eventually gaining access to the repository---a process that usually lasts between two and three years. On the contrary, hired developers experience a sudden integration with an almost simultaneous start-up in mailing lists, bug-tracking system, and repositories. While we also find differences between hired industry developers and open source volunteers, our time frame of interest is much shorter.

The goal of our study is to move from a theoretical understanding of newcomer processes to actionable findings ready to be implemented and evaluated in tool support for project newcomers.

\subsection{Recommender Systems for Newcomers}

Recent advancements in recommender systems for newcomers in software development and open source projects aim to streamline the onboarding process by addressing common barriers such as task identification and project integration. Malheiros et al.~introduced a system to recommend relevant source code, reducing the need for direct mentorship~\cite{malheiros2012source}. Liu et al.~developed NNLRank, a neural network-based system that ranks projects for developers, demonstrating the effectiveness of customized project recommendations~\cite{liu2018recommender}.

Xiao et al.~proposed the Personalized First Issue Recommender (PFIRec), which uses LamdaMART to recommend issues based on individual newcomer profiles, significantly improving task matches and engaging newcomers with suitable tasks~\cite{xiao2023personalized}. In addition, conversational bots have been explored as a means of assisting newcomers. Serrano Alves et al.~presented a chatbot that filters tasks to help newcomers find suitable challenges, indicating that such interfaces can enhance the newcomer experience by simplifying task selection~\cite{serrano2021find}. Similarly, Dominic et al.~proposed a bot that recommends projects and provides resources, to increase newcomer retention through comprehensive support~\cite{dominic2020conversational}.

Furthermore, Fronchetti et al.~investigated the content of CONTRIBUTING files, finding that many do not adequately address common onboarding barriers, highlighting the need for better documentation to facilitate newcomer integration~\cite{fronchetti2023contributing}. Complementing this, Gao et al.~explored the simplification of GitHub README files using transfer learning, aiming to make software documentation more accessible to newcomers. By training a Transformer-based model on a dataset of GitHub README files and fine-tuning it with general-domain data, they significantly improved the comprehensibility of documentation, demonstrating the effectiveness of simplifying technical information to improve the understanding of newcomers~\cite{gao2023evaluating}.

Collectively, these studies emphasize the importance of using technology to provide personalized support and improve the onboarding experience for newcomers to the software development ecosystem.

\section{Research Method}

In this section, we present our research questions and the methods used for data collection and analysis, as well as demographic data about our study participants.

\subsection{Research Questions}

Three research questions guide our research:

\begin{description}
\item[RQ1] What processes do newcomers follow to make their first contribution after entering a project?
\item[RQ2] What information is important to newcomers during this period?
\item[RQ3] How do project and individual characteristics relate to the processes followed by newcomers and the types and sources of information consulted?
\end{description}

\subsection{Data Collection}

\begin{table*}[t]
\begin{center}
\caption{Survey questions (excerpt)}
\label{tab:survey}
\begin{tabular}{ll}
\hline
 & \textit{For the following questions, please consider one specific project that you recently contributed to for the first time}\\
 & \textit{(or tried to). This can be open source or closed source.} \\
1 & Is this project open source or closed source? \textit{(open source/closed source)} \\
2 & What programming language(s) does this project use? \textit{(text box)} \\
3 & Approximately how many newcomers (developers making a code contribution who have never contributed to this \\
 & project before) does this project receive on average? \textit{(drop-down)} \\
4 & What steps did you follow from first hearing about this project to making your first code contribution? \textit{(textbox)} \\
5 & What information sources did you use to obtain the knowledge necessary to make your first code contribution to\\
 & this project? Why? \textit{(textbox)} \\
6 & Do you remember any case of misinformation or lack of information that led to problems? \textit{(textbox)} \\
7 & Imagine that a new person is joining this project now. How would you design a step-by-step guide (including \\
 & information sources) to help them make their first code contribution? \textit{(textbox)} \\
\hline
8 & In your opinion, how important are the following types of information for a newcomer to this project (somebody \\
 & who has already decided to contribute and is trying to place their first code contribution)? \textit{(Likert scales, see} \\
 & \textit{Table~\ref{tab:informationtypes-ratings} for items)} \\
9 & What other types of information are important? \textit{(textbox)} \\
\hline
10 & In your opinion, how important are the following information sources for a newcomer to this project (somebody \\
 & who has already decided to contribute and is trying to place their first code contribution)? \textit{(Likert scales, see} \\
 & \textit{Table~\ref{tab:informationsources-ratings} for items)} \\
11 & What other information sources are important? \textit{(textbox)} \\
\hline
\end{tabular}
\end{center}
\end{table*}

To answer our research questions, we designed a web-based survey. Table~\ref{tab:survey} shows the most important questions of the survey, with each horizontal line representing a page break. The complete survey is available at \url{http://tinyurl.com/NewcomerNeeds}.

Following Fink's advice on survey design~\cite{Fink1995}, we instructed participants to think about a specific scenario---namely the newcomer experience in one of their own projects. We believe that we received more valuable answers asking about the concrete scenario of our participants' own joining experience rather than some hypothetical experience. We did ask participants how long they had been involved with the project, but we did not find any impact of the corresponding answers on the rest of the data. After asking about the characteristics of that project (questions 1-3), we asked participants about the steps they followed when joining the project (question 4) and the ideal step-by-step process (question 7). We also asked about the information sources they needed to get started (question 5) and potential misinformation (question 6). The answer options for the Likert-scale questions about the importance of information types (question 8) and information sources (question 10) were based on Steinmacher et al.'s barriers~\cite{Steinmacher2015} and previous work on API documentation on the web~\cite{Parnin2011}. In addition to the questions in Table~\ref{tab:survey}, we asked about information needs regarding the community and people involved in a project and we collected demographic information from our participants.

To distribute the survey, we contacted the same set of GitHub developers who participated in previous work~\cite{Treude2015}. We randomly sampled these users until we reached saturation in our coding (see next section). We reached saturation after receiving 97 responses. 

To validate our findings, we conducted seven follow-up interviews with a subset of the survey participants. These interviews focused on validating the 16-step model for the processes followed by newcomers and on deepening our insights on information needs and sources. The interviews were conducted over Skype and lasted at least 30 minutes each. The interview script is available at \url{http://tinyurl.com/NewcomerInterviews}.

\subsection{Data Analysis}

Considering the exploratory nature of our first two research questions, we used methods from Grounded Theory~\cite{Charmaz2014} to analyze the collected data, while constantly referring and comparing to related work, in particular to Steinmacher et al.'s barriers model~\cite{Steinmacher2015} and von Krogh et al.'s joining script~\cite{Krogh2003}. We refer to both references in the discussion of our findings. We conducted two collaborative coding sessions of approximately four hours each, involving two of the authors, in which we coded the responses to open-ended questions in the survey. The answers related to the processes and the answers related to the information needs in terms of types and sources were coded separately. We initially coded the responses after receiving about 50 responses but determined that saturation had not been reached. After receiving another 47 responses, we repeated the coding, this time reaching saturation. We obtained 104 codes for the process part of the study and 47 additional codes for the information type and source part of the study. We used axial coding to find higher-level conceptual themes to answer our research questions and formed core themes such as the newcomer processes described in Section~\ref{sec:process}. We grouped interview quotes from validation interviews under the same themes.

To answer our third research question, we quantitatively analyzed the degree to which project and individual characteristics correlated with specific answers to survey questions.

\subsection{Demographics}

Half of the participants in our survey had at least five years of development experience, and many indicated having ten or more years of experience. By far, most of them were active in more than one project. For most of the participants, software development was part of their job: 80 (82\%) of the participants gave a positive response to the corresponding question compared to 15 (15\%) negative responses while 2 participants did not answer the question.

\section{Findings}

In this section, we report our findings on the processes followed by newcomers on their way to their first code contribution (Section~\ref{sec:process}) and on their needs in terms of information types and sources (Section~\ref{sec:information}). For each code that emerged in the qualitative analysis, we indicate how many participants mentioned the particular theme in superscript. Note that these numbers only indicate how much evidence the data analysis yielded for each theme; they do not necessarily indicate the importance of a theme since in qualitative data analysis, frequency does not imply importance and vice versa.

\subsection{Processes}
\label{sec:process}

Figure~\ref{fig:process} depicts the processes of newcomers that emerged from our data. In the following, we describe each step in detail.

The order of steps depends on whether the newcomer has a specific task in mind (e.g., fixing a particular bug in an open source project as a casual contributor~\cite{Pinto2016} or working on an issue assigned to them in an industry project). If so, they commonly start the process by understanding their task in more detail and treat understanding the project as a whole as a secondary goal: \textit{``I understand the problem, researching how best to do, discuss the idea, activities--implement, test and only after that all I share the change''}$_{P6}$. If the newcomer's goal is to become a contributor to a project independent of a specific task, the process starts with setting up the environment, followed by understanding the project and finding the first task to work on: \textit{``Download the code; being able to run the tests; read the main loop to get the feeling of how it works; pick an issue on the issue tracker suitable for a beginner in the project''}$_{P74}$. These alternatives are shown in Figure~\ref{fig:process}. In the following, we describe the process that begins with the setup of the environment. An additional scenario not accounted for in the figure occurs when the developer contributing is the first contributor of the project, i.e., he or she started it. The paths in Figure~\ref{fig:process} are drawn based on ``sequential evidence'', i.e., participants mentioning two activities after each other.

Setting up the environment$^{(12)}$ can be the first step when joining a project: \textit{``First install the project on my machine, that is, the source code of the project''}$_{P19}$. With distributed version control systems, this involves cloning a repository$^{(10)}$: \textit{``I made a git clone and from that I added the changes''}$_{P92}$, and can entail signing up for GitHub$^{(1)}$. This step is completed when the newcomer can run tests$^{(2)}$: \textit{``Get the code and run the tests. Verify that you can make them pass on your local environment''}$_{P74}$. Setting up the environment was also indirectly identified as a newcomer activity by Steinmacher et al.~\cite{Steinmacher2015} who indicated barriers such as ``local environment setup hurdles''. In our study, mentoring was mentioned as a strategy for lowering these barriers: \textit{``The interesting thing would be to have a person to help the beginners: finding bugs, help in the installation of the software''}$_{P19}$, along with proper documentation: \textit{``Scripts to get the repository cloned and set up properly''}$_{P4}$.

After the environment is set up, several steps are needed to get up to speed on the project, its codebase, architecture, process, domain, and related technologies, standards, and techniques. Understanding the project$^{(43)}$ as a whole is important: \textit{``Understand and believe in the project where you invest your time''}$_{P75}$. In this step, developers read documentation$^{(35)}$, often focusing on \textit{readme} and \textit{contributing} files$^{(11)}$: \textit{``First read CONTRIBUTERS.md and README.md to understand the project''}$_{P91}$. Documentation can be automatically generated$^{(1)}$ and can be found in wikis$^{(6)}$ or Read the Docs, tutorials$^{(3)}$, API-style documentation$^{(3)}$, other websites$^{(3)}$, or in emails$^{(1)}$ and cheatsheets$^{(1)}$. Although most of the documentation considered at this stage is in text form, our participants also mentioned workflow diagrams$^{(1)}$ and code demos in video form$^{(2)}$: \textit{``Using a Demo and record it using video''}$_{P94}$. Documentation fulfills many functions and includes information on changing the software as well as on using it$^{(4)}$: \textit{``Write the readme file that contains instructions on how to use, how to compile, how to change the code base and how to submit their changes back''}$_{P79}$.

\begin{figure*}[t]
\centering
\includegraphics[width=\linewidth]{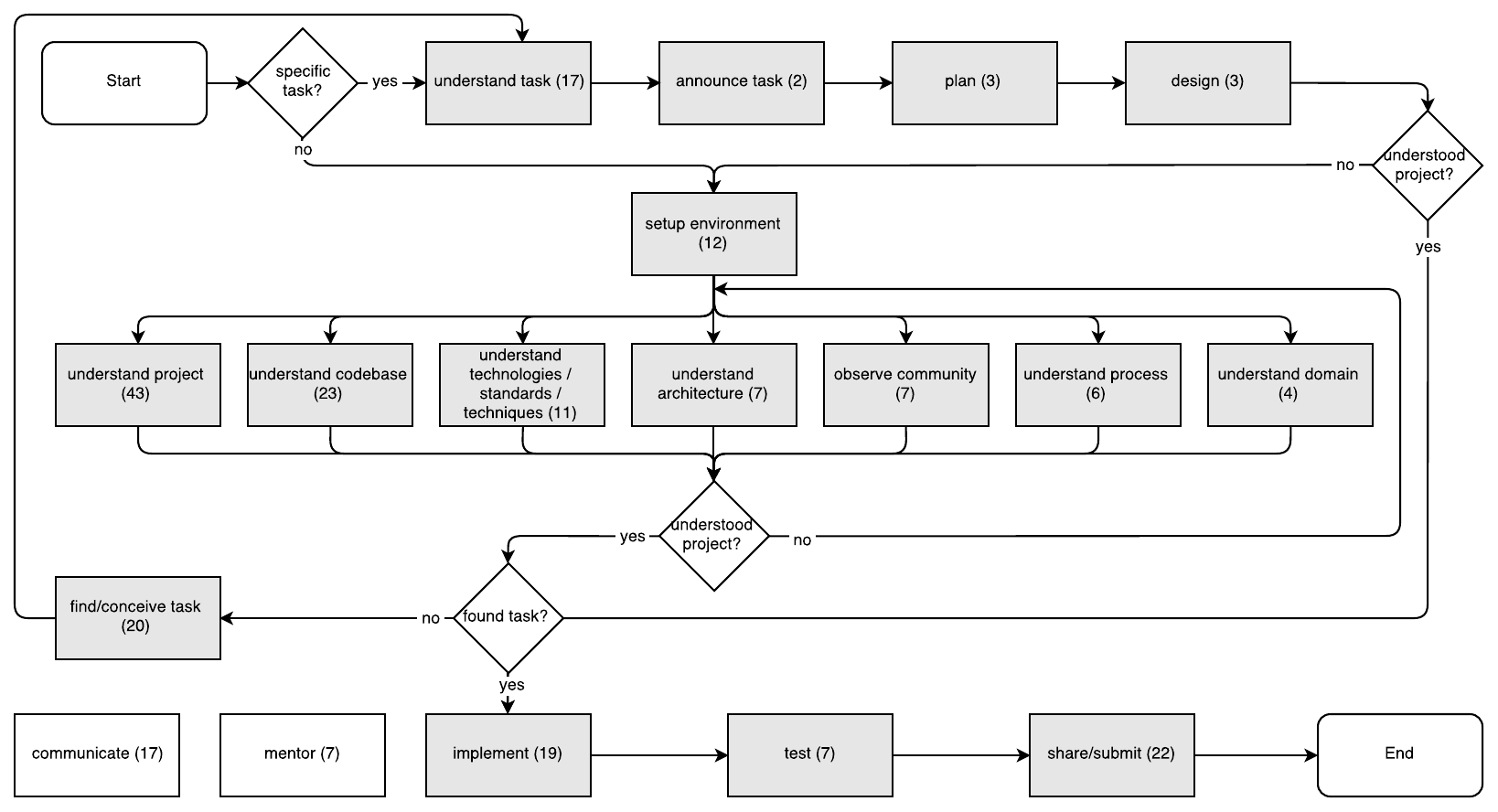}
\caption{Newcomers' processes from after they have decided to contribute until their first code contribution}
\label{fig:process}
\end{figure*}

For newcomers, it is particularly important to understand the goals$^{(8)}$ and the purpose of the project they are joining: \textit{``I would tell them to read the documentation to familiarize themselves with the structure and goal of the project''}$_{P33}$. This involves understanding who the users$^{(4)}$ are and why the project is useful$^{(2)}$: \textit{``A newcomer should know about what and for whom the project is aimed for, the community it will target, the type of end users''}$_{P36}$. Furthermore, it is important for newcomers to understand the responsibilities of other developers on the team and the communication channels used$^{(6)}$: \textit{``I think new developers need access to information about the team and what each member contributes. This way they can best place themselves in the team and also know where to go for help''}$_{P30}$.

The next step is to understand the codebase$^{(23)}$. Tests$^{(5)}$ can be a good starting point for this step: \textit{``Tests are a good way to understand how all the pieces behave and how they communicate to each other''}$_{P3}$, and making small changes$^{(5)}$ is a good way to familiarize oneself with a codebase as a newcomer: \textit{``Read documentation, install the software, try it out, make a few nondramatic changes, create a pull request with my code contribution''}$_{P48}$. Other developers who have modified a given part of the codebase before can be a good source for information$^{(4)}$: \textit{``I'd point them to the right documentation, maybe even to the specific person that developed that part, and Stack Overflow posts''}$_{P89}$. Additional information can be found in source code comments$^{(3)}$, stack traces$^{(1)}$, or documentation$^{(1)}$.

A software project does not exist in isolation, but rather in an ecosystem of technologies, standards, and techniques that a newcomer needs to understand$^{(11)}$: \textit{``Read anything available about the technology used by the project, e.g., on an Android related project: Android development guidelines, Android APIs related to the project on Google's own Android Developers site, as it contains examples and very detailed documentation on APIs and Training Guides on development techniques''}$_{P36}$. Coding style$^{(6)}$ is particularly important as this example illustrates: \textit{``I've seen python code rejected not because it was bad or solved the wrong problem, but because it didn't conform to the formatting and usage standards set on the project, which made it too difficult to integrate''}$_{P93}$. Reading documentation$^{(4)}$ was mentioned as the most common strategy to understand technologies, standards, and techniques relevant to a project.

Understanding the architecture$^{(7)}$ is another important step in the onboarding process of a newcomer. A comprehensive ``Hello World'' example$^{(3)}$ can help explain the architecture to a newcomer: \textit{``Provide them a `Hello World' example program within our architecture''}$_{P13}$. Other sources of information can include papers$^{(1)}$: \textit{``We also have a couple of papers describing the structure of the code and the flow of information''}$_{P27}$, and diagrams$^{(1)}$: \textit{``Instead of text I will provide a graphical diagram to explain the architecture of the project, which can be understood by newcomers and they can dive into their respective part''}$_{P70}$.

In particular in open source projects, a newcomer needs to understand the community around a project and its practices$^{(2)}$. With many of the communication channels available in the open, newcomers can observe the community$^{(7)}$ before interacting with the project actively, as this participant described in his own process of joining an open source project: \textit{``Long-term observation on open source projects on their coding practice''}$_{P58}$. Tools for community observation include mailing lists$^{(3)}$ and IRC channels$^{(2)}$.

Understanding the process$^{(6)}$ followed by a project, in particular the contribution process$^{(5)}$, is a prerequisite for a newcomer to complete their first code contribution successfully: \textit{``Understand the dev `flow' (branching, how to get code pulled, etc)''}$_{P51}$. Git-flow is one way of organizing the contribution process and would need to be explained to newcomers: \textit{``I would then ask them to use the Git-flow model we've set up to create pull requests and incorporate feedback before merges''}$_{P33}$. The importance of understanding the contribution process was also identified in Steinmacher et al.'s barrier model~\cite{Steinmacher2015}, which includes barriers such as ``poor how-to-contribute available'' and ``lack of information on how to send a contribution''. The contribution process also played an important role in our validation interviews: \textit{``What I needed to know was how to do the ceremonies they did, the process stuff. And some of the basics, you have to know how to write software. [...] You might observe that they were doing these things, but without the context and framework of understanding why they're doing it, it might not be comprehensible as to the value of why they're doing things, and then it would be really easy to get them kind of correct, but get fundamental things wrong''}.

Another step in understanding the project landscape that a newcomer is becoming part of is understanding the domain$^{(4)}$ and its terminology$^{(1)}$. In some cases, research$^{(3)}$ is necessary, as this example process mentioned by one participant shows: \textit{``Understand the terminology of the mHealth interdisciplinary domain and how to translate state-of-the-art research into a production and deployable system''}$_{P13}$.

After understanding the project and all related concerns, the next step for a newcomer is to find a task to work on$^{(20)}$. This theme is mirrored in the joining script proposed by von Krogh et al.~\cite{Krogh2003} who refer to it as ``ask for a task to work on''. How to find a task depends on the setting of the newcomer: in open source projects, it is common to use a project$^{(14)}$ and find a bug to fix or a new feature to implement: \textit{``I think code contribution should come from a need for extending functionality or fixing a bug. The first thing I would suggest is: use the existing code, and if you find something else you need, or you find a bug, then get the source code''}$_{P38}$. However, this theme---finding a task to work on by using the project---was exclusively mentioned by participants in open source, and not by participants who worked on industry projects where it is more common for tasks to be assigned by somebody on the team. For newcomers, focusing on simple and/or low-priority tasks$^{(6)}$ can make it easier to get up to speed: \textit{``Take a look at the issues. Pick one that seems easy enough to be fixed. Some projects have a label such as `help wanted' or `good first contribution', so you may want to look at it''}$_{P74}$. Mentoring$^{(2)}$ can help in finding a good task to start with.

The next step after finding a task is understanding it$^{(17)}$, an activity that can be supported by source code$^{(6)}$, issue trackers$^{(3)}$, Stack Overflow$^{(3)}$, search engines$^{(2)}$, and attempts to reproduce behavior$^{(2)}$: \textit{``If something doesn't work, seek help from the GitHub site, search engine and Stack Overflow. If you cannot get it work, examine the source code to see where does it go wrong''}$_{P20}$. This is also the phase in which community feedback$^{(2)}$ on the task can be obtained using a prototype$^{(1)}$: \textit{``I did some prototypes to demonstrate the desired features, then some local tests within a local repository and finally created a shared one where I've uploaded the code''}$_{P44}$. Some of our participants mentioned following a requirements engineering$^{(2)}$ process to support their understanding of the task: \textit{``I just followed the requirement engineering like the IEEE does''}$_{P11}$. A crucial part of understanding a task is understanding its completion criteria$^{(2)}$, i.e. when the task is completed: \textit{``Most of the information required is kind of `how can I say that this issue is resolved'. If you start coding away before answering this question it will definitely have issues when submitting your code to review. I've seen this happen where the issue opened was too vague and not even the one who opened it had a clear idea on how to specify `this issue is done' ''}$_{P74}$. In addition, the participants mentioned the importance of understanding the value that would be added$^{(1)}$ to a project by completing a task: \textit{``1. Does your contribution add value to the overall project? 2. Does your contribution ease the use of the project?''}$_{P69}$.

For newcomers to open source projects, announcing the task that they intend to work on$^{(2)}$ can be a step to avoid duplicate work. This is usually achieved through the creation of a new issue in the project's issue tracker$^{(2)}$: \textit{``They need to first open an issue about the new feature or change they want to contribute before just making a new pull request''}$_{P50}$. This step was also identified in von Krogh et al.'s work~\cite{Krogh2003} as ``announcing `external' contribution'' and ``propose/outline bugfix (no code)''.

Depending on the size of the task, planning$^{(3)}$ is required to determine how a task fits into the rest of the project. The importance of project management and a project lead$^{(2)}$ was emphasized by our participants: \textit{``Remember that there is a main author who's got a plan for the project so don't pull it in your own direction''}$_{P35}$.

The last step before implementation is the design$^{(3)}$, again depending on the size and nature of the task: \textit{``First use the mailing list, ask about the bug or talk about the design if this is a new feature or enhancement''}$_{P67}$. A sketch$^{(1)}$ can be used to communicate the design to other stakeholders in the project: \textit{``I met the clients and discussed with the whole team. I created sketches of how I think it should work''}$_{P30}$.

Once the task and the project are understood, newcomers can start implementation$^{(19)}$. Important considerations during this phase include the coding style$^{(5)}$: \textit{``Work on the code and go back to doc or mailing list for questions. Check previous code contribution to see how it must be done in terms of code (syntax, naming, etc...)''}$_{P67}$, as well as testing style$^{(3)}$, as this example from one of our participants shows: \textit{``[I would] familiarize myself with their flavor of TDD (tooling, process, etc)''}$_{P51}$.

After implementation, the next step is testing$^{(7)}$: \textit{``Then write tests first with expected behavior, then make changes to fix failing tests''}$_{P91}$. The tests should first be run on the local machine$^{(2)}$.

The last step in the newcomer's processes is sharing or submitting the produced code$^{(22)}$. Especially in open source projects, newcomers are encouraged to do just that: \textit{``Don't be afraid of submitting your code, and don't be afraid of comments. Having feedback on your pull request is the norm rather than the exception''}$_{P74}$. The details of the submission process depend on the version control system and the project's practices, but the following quote is representative of many of our participants: \textit{``If your fix works, fork the project to your account on GitHub. Make a separate branch for the fix in your own fork. Put the fix to the branch. Try that you can use the fixed version from your own branch. Once it works, make a pull request from the branch''}$_{P20}$. Pull requests$^{(11)}$, forks$^{(3)}$, and branches$^{(3)}$ were all mentioned by several of our participants. While technically not part of submitting the first code contribution, our participants also mentioned the review process, in particular in terms of the patience$^{(1)}$ required by newcomers: \textit{``There is a quality process to be followed before merging the code in the mainline. This review is what grants the project quality and that the developer didn't make any mistake''}$_{P18}$.

Two additional activities that emerged from our analysis were mentioned in conjunction with many of the steps outlined above, and we present them here as cross-cutting strategies: communication$^{(17)}$ and mentoring$^{(7)}$. Communication is essential to coordinate a newcomer's journey towards their first code contribution, and in particular in open source projects, communication can help lower the entry barriers for newcomers: \textit{``Mostly everyone I have met has been willing to help, you just need to ask, demonstrate that you are trying your best and willing to listen to them''}$_{P72}$. Communication also plays an important role in von Krogh et al.'s joining script~\cite{Krogh2003} with steps such as ``coordination and organization discussion'', ``off-topic discussion'', and ``discuss legal/philosophical implications/matters''.

Mentoring plays a cross-cutting role as well and has been adopted by many projects: \textit{``Whenever someone new arrives, we instruct someone older in the project to be your mentor and accompany him for a while, until the new person has basic knowledge about the project''}$_{P6}$, and pair programming$^{(1)}$ can be one way of implementing a mentoring strategy: \textit{``I can join in a pair programming session to help them with their commit''}$_{P48}$.

In addition to the steps outlined above, our participants mentioned barriers$^{(17)}$ that they encountered during their attempts to contribute their first code. We cross-referenced these barriers with those identified by Steinmacher et al.~\cite{Steinmacher2015} and were able to confirm many of their barriers, but did not find new ones. 

All participants we talked to during follow-up interviews validated the model. One interviewee stated: \textit{``I can relate to my thought process going through the chart, it definitely makes sense. [...] The communication part [...] is kind of interwoven through a lot of the different steps, I agree with it being on the side, it doesn't necessarily have a particular place in the chart, it's kind of an overarching component in the process.''} Another interviewee explained the other cross-cutting strategy of mentoring: \textit{``I like the mentor part of it. [...] Because then you don't get stuck in this middle section [of the diagram]. [...] That middle section is where you get caught a lot of the time, you just never leave it''}. Yet another interviewee confirmed: \textit{``The flow seems very pertinent. The guy has to understand the task, advertise to people that he is going to do that, the community discusses a plan with the developer, then you provide your code''}.

In the validation interviews, we explicitly asked whether the newcomer process should be different based on open vs.~closed source: \textit{``For me it wasn't, the one that I contributed to, I had a specific task in mind and I just did the thing, so maybe not''}. Another interviewee added: \textit{``I would say industry is slightly---it's going to be a lot easier. Because most of the time you're doing a specific task''}. Other interviewees echoed these opinions. Thus, we decided not to separate the model into one for industry and one for open source, but acknowledge that starting the newcomer process with a specific task is more common in industry projects.

\subsection{Information Sources and Types}
\label{sec:information}

In this section, we detail our findings on the information sources and types that are important to newcomers during this period.

\subsubsection{Information Sources}
\label{sec:information-sources}

Many of our participants mentioned sources$^{(69)}$ of information that they had considered when joining a project themselves or that they would recommend to other newcomers joining their project.

First, different types of websites$^{(20)}$ are an important source of information for newcomers joining a project: \textit{``Various online web development websites. Almost all web development resources have to be current. Print resources are too stale''}$_{P24}$. Particularly in industry settings, other developers$^{(18)}$ are an essential source of information: \textit{``Other developers, because they were the only source of that knowledge''}$_{P4}$. This theme---obtaining information from other developers---was mentioned significantly more by participants in industry settings than by participants in an open source setting. Documentation$^{(16)}$, in general, helps newcomers get up to speed: \textit{``I used the project's documentation and the language guide to understand core functions used''}$_{P17}$, and search engines$^{(14)}$ are an effective way for making sense of documentation: \textit{``Googling since I've become a pro after narrowing down my searches''}$_{P72}$. Related to search engines is the use of the Question and Answer website Stack Overflow$^{(10)}$: \textit{``I used Google and Stack Overflow because I have used them before with good results''}$_{P84}$. Participants mentioned using their own experience or education$^{(13)}$ while making their first code contribution: \textit{``I leveraged my knowledge from college and experience developing hardware as a contractor''}$_{P23}$. Education can happen online as this example shows: \textit{``I use classes of schoolofnet.com because it was easy to learn from there''}$_{P14}$. In addition, the source code$^{(13)}$ plays a central role in making sense of a new project: \textit{``The code should be really self-explanatory. If it isn't, then people won't contribute, pure and simple. People don't contribute to code that is badly written, doesn't perform, and doesn't have some sort of documentation, because nobody uses it''}$_{P38}$. To understand the code, debugging$^{(4)}$ can be used: \textit{``Debugging, mostly to understand what was broken, what was working and what was doing unnecessary work under the hood''}$_{P36}$. Tests$^{(6)}$ are part of source code and can play an important role in aiding a newcomer's understanding: \textit{``Also, I believe that if you are writing code test first, the tests are largely all the documentation you'll need after being given an overview of the architecture''}$_{P51}$. Similar to the role of other developers, this theme---using tests as a source of information---was mentioned more often by participants in industry projects than by those in open source projects. GitHub pull requests$^{(3)}$ can be an important source for understanding the contribution process of a project: \textit{``Contribution guide as shown in README and Pull Requests to generally understand how other people contribute their code, how did the original fork accept contributions, and what are the basic steps on contributing other projects to understand what are the common contribution practices/coding practices in each project''}$_{P58}$. This theme---using pull requests to understand a project's contribution process---confirms the finding of related work~\cite{Gharehyazie2013} regarding the importance of patch activity in open source projects. Other sources that were mentioned include forums$^{(5)}$, Wikipedia$^{(2)}$, books$^{(1)}$, commits$^{(1)}$, issue trackers$^{(1)}$, papers$^{(1)}$, and videos$^{(1)}$.

\begin{table}
\begin{center}
\caption{Rating of information sources}
\label{tab:informationsources-ratings}
\begin{tabular}{lrr}
\hline
\textbf{information source} & \textbf{mean} & \textbf{correlation with activity} \\
\hline
Source Code & 4.60 & -0.08    \\ 
API Documentation & 4.43 & 0.15 \\   
Issue Trackers & 3.94 & -0.02  \\
Mentors & 3.76 & -0.07 \\
Tutorials & 3.70 & *-0.31 \\  
Stack Overflow & 3.61 & 0.02\\   
Forums & 3.31 & -0.24   \\
Blogs & 3.20  & -0.09  \\
Mailing Lists & 3.08 & -0.10\\   
\hline
\end{tabular}
\end{center}
\end{table}

The two columns on the left of Table~\ref{tab:informationsources-ratings} show the average rating that our participants assigned to different information sources on a Likert scale. The most important item is source code with a rating of 4.60 on a scale from 1 (not important at all) to 5 (very important), followed by API documentation. Forums, blogs, and mailing lists were considered the least important among the answer options, yet still with a slightly higher than neutral rating. We hypothesize that the perceived importance of different information sources depends on the context of the newcomer, both in terms of personal characteristics and characteristics of the project to which they are making a code contribution. To investigate this hypothesis, we analyzed the correlation between the characteristics of our participants (such as their activity level on GitHub) and the answers they gave, as well as the correlation between project characteristics (such as the extent of newcomer influx or programming languages) and their answers.

The right column of Table~\ref{tab:informationsources-ratings} shows the Spearman correlations between the ratings for the different information sources and the GitHub activity of our participants (public contributions, i.e., push, pull, or issue creation) measured in terms of the number of contributions over the last 12 months. We hypothesize that developers with more activity perceive different sources as more important compared to those with less activity. As the right column of Table~\ref{tab:informationsources-ratings} shows, almost all correlations are negative, except API documentation and Stack Overflow, suggesting that more active developers value these two sources more while they perceive other sources as less important. One of the correlations is statistically significant (assuming 9 hypotheses following the Benjamini-Hochberg procedure~\cite{Benjamini1995} to account for multiple comparisons and an alpha level of .05): Tutorials are seen as significantly less important by participants with more activity on GitHub. We speculate that tutorials are aimed at less experienced developers, and that developers with more activity need more specialized information. Future work is needed to investigate this further.

\begin{table}
\begin{center}
\caption{Impact of newcomer influx on ratings}
\label{tab:informationsources-influx}
\begin{tabular}{lrr}
\hline
\textbf{type of information} & \multicolumn{2}{c}{\textbf{newcomers/year}} \\
 & \textbf{$<2$} & \textbf{$\geq2$} \\
\hline
Source Code & 4.51 & 4.78 \\
API Documentation & 4.43 & 4.41 \\
Issue Trackers & 3.81 & 4.11 \\
Mentors & 3.78 & 3.74 \\
Tutorials & 3.89 & 3.41 \\
Stack Overflow & 3.79 & 3.30 \\
Forums & 3.35 & 3.16 \\
Blogs & 3.29 & 3.08 \\
Mailing Lists & 2.90 & 3.25 \\
\hline
\end{tabular}
\end{center}
\end{table}

Table~\ref{tab:informationsources-influx} shows the average rating for information sources, separately for projects with fewer than two newcomers per year (54 projects in our data) and projects with at least two newcomers per year (37 projects in our data). Although there are some differences in the means (e.g., tutorials, Stack Overflow, forums, and blogs all received a higher rating for projects with few newcomers, while source code, issue trackers, and mailing lists were rated higher in the context of projects with high newcomer influx), none of them is statistically significant (Mann-Whitney-Wilcoxon test, assuming 9 hypotheses following the Benjamini-Hochberg procedure~\cite{Benjamini1995} and an alpha level of .05). 

When we analyzed the impact of a project's programming language on the rating of information sources, one difference was statistically significant (following the Benjamini-Hochberg procedure~\cite{Benjamini1995} and an alpha level of .05): mailing lists were seen as significantly more important for projects that use C (mean: 4.30) compared to those projects that do not use C (mean: 2.90). We compared the ratings for the five most frequently used programming languages in the projects of our participants: Java, JavaScript, Python, C, and PHP. This difference is another piece of evidence suggesting that tool support for newcomers has to take into account project and newcomer characteristics.

\subsubsection{Types of Information}

\begin{table}
\begin{center}
\caption{Rating of types of information}
\label{tab:informationtypes-ratings}
\begin{tabular}{lrr}
\hline
\textbf{type of information} & \textbf{mean}  & \textbf{correlation} \\
 & & \textbf{with activity} \\
\hline
tools and technologies used & 4.15 & -0.14 \\  
programming languages used & 4.15 & -0.09   \\
arch., design, project structure & 4.05 & -0.12 \\  
project domain & 4.02 & 0.02   \\
platform and library dependencies & 4.00 & -0.13    \\
how to set up \& build workspace & 3.93 & -0.22  \\ 
project process and practices & 3.90 & -0.03   \\
how to make a code contribution & 3.89 & -0.23   \\
finding artifacts to modify & 3.73 & 0.10  \\   
finding good tasks for newcomers & 3.52 & -0.27   \\
finding a mentor & 3.41 & -0.14    \\
project culture & 3.24 & -0.16  \\
\hline
\end{tabular}
\end{center}
\end{table}

We investigated the types of information relevant to a newcomer from the perspective of the barriers identified in previous work~\cite{Steinmacher2015} and the misinformation received by our participants. The two columns on the left of Table~\ref{tab:informationtypes-ratings} show the rating of the participants for different types of information. The information on the tools, technologies, and programming languages used was seen as most important, while the information on how to find a mentor and the information on the project culture was seen as less important, although still receiving a higher than neutral rating on average. 

Our participants mentioned some additional types of information$^{(20)}$ in the open-ended part of the survey. For example, information on the coding style$^{(6)}$ used by a project is important: \textit{``The style of code to follow, such as spacing between parentheses in certain places, placement of brackets, and if statements always having brackets''}$_{P84}$. Also, information on the contribution process$^{(6)}$ plays an important role: \textit{``Information about the workflow are very important. How does the workflow work? How to deploy, how to work on tasks and which standards or conventions does the team have?''}$_{P11}$. Another piece of information that is relevant to newcomers is how to ask questions$^{(3)}$: \textit{``We mostly care about how well you are able to figure out what's going on, and how you can communicate your trouble or concerns. We're open and willing to help, so long as you ask for it''}$_{P33}$. In addition, the following types of information emerged from our study: legal$^{(2)}$, API usage$^{(1)}$, how to test$^{(1)}$, support and maintenance$^{(1)}$, terminology$^{(1)}$, and users$^{(1)}$.

Similar to the analysis described in Section~\ref{sec:information-sources}, we investigated the impact of individual and project characteristics on the perceived importance of different types of information. The right column of Table~\ref{tab:informationtypes-ratings} shows the correlations between the perceived importance of information types and the activity level of the participants in terms of the number of contributions. The majority of correlations are negative, suggesting that these types of information are less important for more active developers. However, since none of the correlations is statistically significant, we will revisit this topic in more detail in future work.

\begin{table}
\begin{center}
\caption{Impact of newcomer influx on ratings}
\label{tab:informationtypes-influx}
\begin{tabular}{lrr}
\hline
\textbf{type of information} & \multicolumn{2}{c}{\textbf{newcomers/year}} \\
 & \textbf{$<2$} & \textbf{$\geq2$} \\
\hline
tools and technologies used* & 4.41 & 3.77 \\
programming languages used* & 4.46 & 3.63 \\
arch., design, project structure & 4.11 & 4.00 \\
project domain & 3.90 & 4.23 \\
platform and library dependencies & 4.11 & 3.80 \\
how to set up \& build workspace & 4.04 & 3.77 \\
project process and practices & 4.00 & 3.79 \\
how to make a code contribution & 3.98 & 3.77 \\
finding artifacts to modify & 3.60 & 3.97 \\
finding good tasks for newcomers & 3.52 & 3.50 \\
finding a mentor & 3.43 & 3.38 \\
project culture & 3.38 & 3.09 \\
\hline
\end{tabular}
\end{center}
\end{table}

Table~\ref{tab:informationtypes-influx} shows the perceived importance of information types separately for projects with fewer than two newcomers per year and projects with at least two newcomers in the same time frame. With two exceptions (project domain and finding artifacts to modify), the perceived importance of all types of information is higher for projects with few newcomers. Two differences are statistically significant, assuming 12 hypotheses following the Benjamini-Hochberg procedure~\cite{Benjamini1995} and applying the Mann-Whitney-Wilcoxon test at an alpha level of 0.05: Information on tools, technologies, and programming languages used was seen as significantly more important in projects with few newcomers. We hypothesize that projects with a high newcomer influx are more likely to employ commonly-used tools, technologies, and programming languages that require less specialized information to be communicated to newcomers. Future work should investigate this hypothesis.

Finally, our analysis of the impact of programming languages on the perceived importance of different types of information did not reveal statistically significant results.

\section{Practical Implications}

Our ultimate goal is to automatically identify, extract, generate, summarize, and present documentation that is relevant to newcomers. As a first step towards this goal, we need to understand the steps that newcomers undertake to make their first code contribution. Previous models of newcomer processes lack empirical validation and are often not detailed enough to be amenable to tool support. For example, FLOSScoach's first step---``Learn about the project and skills''~\cite{Steinmacher2016}---is generic and needs to be supported by a range of tools rather than a single tool. In this work, we built a model of newcomer processes based on empirical data, and were able to capture steps at a more detailed level. For example, ``Learn about the project and skills'' corresponds to at least seven distinct steps in our model. By detailing the steps, researchers and practitioners may come up with new ideas and tools to tackle each of them, fine-tuning them according to specific needs.

Tool support for newcomers should take into account the steps identified in our model and support them, primarily with documentation about how each step is typically achieved. The most encouraging finding of our study is that the types of information that received the highest rating from our participants (information on tools, technologies, and programming languages used) are relatively easy to identify and communicate to newcomers. Tool support that automatically analyzes a given project, determines the tools, technologies, and programming languages used by the project, and points newcomers to learning resources for these is not impossible to build, e.g., using generative AI~\cite{xiao2024devgpt}.

However, simply invoking generative AI as a catch-all solution overlooks the nuanced understanding and contextual insights required to effectively support newcomers. Our model of newcomer processes, derived from empirical data, could play a critical role in shaping the design of prompts and resources. By integrating our model, AI tools can be guided to generate documentation and tutorials that are not only technically accurate, but also aligned with the sequential steps and specific challenges identified for newcomers. This approach ensures that AI-generated content is highly relevant and addresses the precise needs and questions that newcomers have at each stage of their onboarding journey.

Furthermore, by using our model, we can instruct generative AI systems to account for the diversity of learning styles, prior knowledge, and personal goals of newcomers. This tailored approach goes beyond generic tutorials and documentation, creating a learning experience that is adaptable and responsive to the individual's progress and feedback, similar to on-demand developer documentation~\cite{robillard2017demand}. In this way, our model not only informs the content of what is generated but also the pedagogical approach, ensuring that the AI support provided is genuinely effective in fostering a welcoming and empowering environment for newcomers.

Similarly, there are plenty of tools that automatically determine the architecture of a project, its platform, and library dependencies. As many of our participants suggested, the contribution process of a project can be understood by analyzing the contribution flow of other developers, for example, in the form of pull requests. All of these information types were highly rated by our participants in terms of importance, whereas the type that would arguably be the hardest to capture automatically---project culture---received the lowest rating.

Another important finding is that individual and project characteristics influence the types and sources of information that are considered important. In particular, our findings suggest that the newcomer support needed depends on how many other newcomers a project has and that the sources to be considered depend on the channels that developers of certain technologies use, as well as on the experience level of the newcomer. It is possible to determine these parameters automatically (e.g., by calculating the number of newcomers to a project in a particular time frame, by identifying the programming language of a project, or by analyzing the previous activity of a developer on other projects) and to customize newcomer documentation accordingly. Building a prototype of this ``parametrized documentation portal for newcomers'' is the next step in our work.

Participants in our validation interviews also emphasized the need for such a tool: \textit{``One cohesive statement saying here's exactly what you need to do to contribute to a software project, I don't know that I've seen that anywhere''}. Another participant added: \textit{``I would definitely say for newcomers it's better to have a very defined starting point, that would definitely help''}. Summarizing and restructuring read-me documentation on GitHub could be a good starting point: \textit{``I'd be very interested to see how well you do with that when you start hitting GitHub. A lot of the starting documentation is on a read-me, but there's no set structure. I've seen so many different ways to do the documentation''}. The cloud-based team collaboration tool Slack~\cite{Lin2016} could be another useful starting point: \textit{``If someone were to go into Slack and Google some words related to an issue that they have, it's probably in the history somewhere with context why the conversation was there in the first place''}.

\section{Limitations}

We chose methods from Grounded Theory to answer our first two research questions due to the exploratory nature of these questions. Although we achieved saturation when analyzing the survey responses, we cannot claim that we recorded all possible perspectives on these questions among software developers. However, we were able to validate the 16-step model through our follow-up interviews.

Our data is imbalanced towards open source projects: our dataset includes 26 closed source projects and 70 open source projects. To mitigate this threat, we explicitly asked all participants in validation interviews about differences in the newcomers' processes and information needs between open and closed source, but we were not able to identify any differences significant enough to invalidate our model.

As mentioned above, while we report the amount of evidence for each theme yielded by qualitative data analysis, we cannot infer the importance of a theme from these numbers alone.

\section{Conclusions and Future Work}

As a step towards supporting newcomers in overcoming the many entry barriers to a software project, we presented the results of an empirical study aimed at identifying the processes and information needs of newcomers. Based on a survey with about 100 practitioners and validation interviews, we developed a 16-step model for the processes that newcomers follow, and we identified their information needs along with the sources they consider. We were able to confirm that much of the information needed by newcomers already exists in some form, but that additional tool support is needed for conveying this information in a form that is easily accessible by newcomers. Recognizing the diversity among newcomers, we noted that a one-size-fits-all solution is impractical, highlighting the importance of tailoring support to individual and project-specific characteristics.

Our future work is focused on the development of the tool support envisioned by our participants: a parametrized documentation portal. This portal aims to automatically extract and generate information relevant to a newcomer from the vast amount of information available to today's software developers, presenting it in a summarized and easily digestible form. The integration of generative AI into this process is a key aspect of our future plans. By leveraging generative AI, we aim to enhance the portal's ability to dynamically generate and tailor documentation and learning resources to fit the specific needs of newcomers. This involves not just summarizing existing information but also creating new content where gaps are identified, ensuring that newcomers have access to comprehensive and up-to-date resources that are directly relevant to their learning path and project involvement.

Moreover, the use of generative AI will allow for a more nuanced understanding of the newcomer's experience level and preferences, enabling the portal to adapt the complexity and presentation of the information accordingly. This personalized approach will help bridge the gap between the information that is currently available and the specific needs of each newcomer, facilitating a smoother transition into the project and increasing the likelihood of a successful first contribution.

Our continued efforts will be directed toward implementing and refining this AI-enhanced documentation portal, with the ultimate goal of providing targeted, accessible, and effective support to newcomers to a software development project.

\end{sloppy}

\begin{thebibliography}{10}

\bibitem{Begel2008}
A.~Begel and B.~Simon.
\newblock Novice software developers, all over again.
\newblock In {\em Proceedings of the Fourth International Workshop on Computing Education Research}, pages 3--14, 2008.

\bibitem{Benjamini1995}
Y.~Benjamini and Y.~Hochberg.
\newblock Controlling the false discovery rate: {A} practical and powerful approach to multiple testing.
\newblock {\em Journal of the Royal Statistical Society. Series B (Methodological)}, 57(1):289--300, 1995.

\bibitem{bradley2022sources}
N.~C. Bradley, T.~Fritz, and R.~Holmes.
\newblock Sources of software development task friction.
\newblock {\em Empirical Software Engineering}, 27(7):175, 2022.

\bibitem{Burnett2014}
M.~M. Burnett and B.~A. Myers.
\newblock Future of end-user software engineering: Beyond the silos.
\newblock In {\em Proceedings of the FSE/SDP Workshop on Future of Software Engineering}, pages 201--211, 2014.

\bibitem{Charmaz2014}
K.~Charmaz.
\newblock {\em Constructing grounded theory}.
\newblock Sage Publications Inc., 2014.

\bibitem{Crowston2008}
K.~Crowston, K.~Wei, J.~Howison, and A.~Wiggins.
\newblock Free/libre open-source software development: What we know and what we do not know.
\newblock {\em ACM Computing Surveys}, 44(2):7:1--7:35, 2008.

\bibitem{Czarnecki2012}
K.~Czarnecki, Z.~Malik, and R.~Lotufo.
\newblock Modelling the hurried bug report reading process to summarize bug reports.
\newblock In {\em Proceedings of the International Conference on Software Maintenance}, pages 430--439, 2012.

\bibitem{Dagenais2010}
B.~Dagenais, H.~Ossher, R.~K.~E. Bellamy, M.~P. Robillard, and J.~P. de~Vries.
\newblock Moving into a new software project landscape.
\newblock In {\em Proceedings of the 32nd International Conference on Software Engineering - Volume 1}, pages 275--284, 2010.

\bibitem{dominic2020conversational}
J.~Dominic, J.~Houser, I.~Steinmacher, C.~Ritter, and P.~Rodeghero.
\newblock Conversational bot for newcomers onboarding to open source projects.
\newblock In {\em Proceedings of the IEEE/ACM 42nd International Conference on Software Engineering Workshops}, pages 46--50, 2020.

\bibitem{fang2020need}
H.~Fang, D.~Klug, H.~Lamba, J.~Herbsleb, and B.~Vasilescu.
\newblock Need for tweet: How open source developers talk about their github work on twitter.
\newblock In {\em Proceedings of the 17th International Conference on Mining Software Repositories}, pages 322--326, 2020.

\bibitem{Fink1995}
A.~Fink.
\newblock {\em How to Ask Survey Questions}.
\newblock Sage Publications Inc., 1995.

\bibitem{Fleming2013}
S.~D. Fleming, C.~Scaffidi, D.~Piorkowski, M.~Burnett, R.~Bellamy, J.~Lawrance, and I.~Kwan.
\newblock An information foraging theory perspective on tools for debugging, refactoring, and reuse tasks.
\newblock {\em ACM Transactions on Software Engineering and Methodology}, 22(2):14:1--14:41, 2013.

\bibitem{Fritz2010}
T.~Fritz and G.~C. Murphy.
\newblock Using information fragments to answer the questions developers ask.
\newblock In {\em Proceedings of the 32nd International Conference on Software Engineering - Volume 1}, pages 175--184, 2010.

\bibitem{fronchetti2023contributing}
F.~Fronchetti, D.~C. Shepherd, I.~Wiese, C.~Treude, M.~A. Gerosa, and I.~Steinmacher.
\newblock Do contributing files provide information about oss newcomers’ onboarding barriers?
\newblock In {\em Proceedings of the 31st ACM Joint European Software Engineering Conference and Symposium on the Foundations of Software Engineering}, pages 16--28, 2023.

\bibitem{gao2023evaluating}
H.~Gao, C.~Treude, and M.~Zahedi.
\newblock Evaluating transfer learning for simplifying github readmes.
\newblock In {\em Proceedings of the 31st ACM Joint European Software Engineering Conference and Symposium on the Foundations of Software Engineering}, pages 1548--1560, 2023.

\bibitem{geng2023interpretation}
M.~Geng, S.~Wang, D.~Dong, H.~Wang, S.~Cao, K.~Zhang, and Z.~Jin.
\newblock Interpretation-based code summarization.
\newblock In {\em Proceedings of the 31st IEEE/ACM International Conference on Program Comprehension}, 2023.

\bibitem{gerosa2021shifting}
M.~Gerosa, I.~Wiese, B.~Trinkenreich, G.~Link, G.~Robles, C.~Treude, I.~Steinmacher, and A.~Sarma.
\newblock The shifting sands of motivation: Revisiting what drives contributors in open source.
\newblock In {\em Proceedings of the IEEE/ACM 43rd International Conference on Software Engineering (ICSE)}, pages 1046--1058. IEEE, 2021.

\bibitem{Gharehyazie2013}
M.~Gharehyazie, D.~Posnett, and V.~Filkov.
\newblock Social activities rival patch submission for prediction of developer initiation in {OSS} projects.
\newblock In {\em Proceedings of the 29th International Conference on Software Maintenance}, pages 340--349, 2013.

\bibitem{gregory2020onboarding}
P.~Gregory, D.~E. Strode, R.~AlQaisi, H.~Sharp, and L.~Barroca.
\newblock Onboarding: How newcomers integrate into an agile project team.
\newblock In {\em Proceedings of the International conference on agile software development}, pages 20--36. Springer International Publishing Cham, 2020.

\bibitem{Haiduc2010}
S.~Haiduc, J.~Aponte, L.~Moreno, and A.~Marcus.
\newblock On the use of automated text summarization techniques for summarizing source code.
\newblock In {\em Proceedings of the 17th Working Conference on Reverse Engineering}, pages 35--44, 2010.

\bibitem{han2024understanding}
J.~Han, J.~Zhang, D.~Lo, X.~Xia, S.~Deng, and M.~Wu.
\newblock Understanding newcomers’ onboarding process in deep learning projects.
\newblock {\em IEEE Transactions on Software Engineering}, 2024.

\bibitem{haque2022semantic}
S.~Haque, Z.~Eberhart, A.~Bansal, and C.~McMillan.
\newblock Semantic similarity metrics for evaluating source code summarization.
\newblock In {\em Proceedings of the 30th IEEE/ACM International Conference on Program Comprehension}, pages 36--47, 2022.

\bibitem{Herraiz2006}
I.~Herraiz, G.~Robles, J.~J. Amor, T.~Romera, and J.~M. Gonz\'{a}lez~Barahona.
\newblock The processes of joining in global distributed software projects.
\newblock In {\em Proceedings of the International Workshop on Global Software Development for the Practitioner}, pages 27--33, 2006.

\bibitem{Jergensen2011}
C.~Jergensen, A.~Sarma, and P.~Wagstrom.
\newblock The onion patch: Migration in open source ecosystems.
\newblock In {\em Proceedings of the 19th Symposium and the 13th European Conference on Foundations of Software Engineering}, pages 70--80, 2011.

\bibitem{kamienski2023analyzing}
A.~Kamienski, A.~Hindle, and C.-P. Bezemer.
\newblock Analyzing techniques for duplicate question detection on q\&a websites for game developers.
\newblock {\em Empirical Software Engineering}, 28(1):17, 2023.

\bibitem{Ko2007}
A.~J. Ko, R.~DeLine, and G.~Venolia.
\newblock Information needs in collocated software development teams.
\newblock In {\em Proceedings of the 29th International Conference on Software Engineering}, pages 344--353, 2007.

\bibitem{koh2021bug}
Y.~Koh, S.~Kang, and S.~Lee.
\newblock Bug report summarization using believability score and text ranking.
\newblock In {\em Proceedings of the international conference on artificial intelligence in information and communication (ICAIIC)}, pages 117--120. IEEE, 2021.

\bibitem{koswatte2021optimized}
D.~D. Koswatte and S.~Hettiarachchi.
\newblock Optimized duplicate question detection in programming community q\&a platforms using semantic hashing.
\newblock In {\em Proceedings of the 10th International Conference on Information and Automation for Sustainability (ICIAfS)}, pages 375--380. IEEE, 2021.

\bibitem{kuttal2021end}
S.~K. Kuttal, S.~Y. Kim, C.~Martos, and A.~Bejarano.
\newblock How end-user programmers forage in online repositories? an information foraging perspective.
\newblock {\em Journal of Computer Languages}, 62:101010, 2021.

\bibitem{Labuschagne2015}
A.~Labuschagne and R.~Holmes.
\newblock Do onboarding programs work?
\newblock In {\em Proceedings of the 12th Working Conference on Mining Software Repositories}, pages 381--385, 2015.

\bibitem{LaToza2010}
T.~D. LaToza and B.~A. Myers.
\newblock Hard-to-answer questions about code.
\newblock In {\em Proceedings of the Workshop on the Evaluation and Usability of Programming Languages and Tools}, pages 8:1--8:6, 2010.

\bibitem{Lawrance2013}
J.~Lawrance, C.~Bogart, M.~Burnett, R.~Bellamy, K.~Rector, and S.~D. Fleming.
\newblock How programmers debug, revisited: An information foraging theory perspective.
\newblock {\em IEEE Transactions on Software Engineering}, 39(2):197--215, 2013.

\bibitem{leon2023comparing}
S.~Leon, M.~Tamanna, and S.~K. Kuttal.
\newblock Comparing foraging behavior across code hosting and q\&a platforms through a gender lens.
\newblock In {\em Proceedings of the IEEE Symposium on Visual Languages and Human-Centric Computing (VL/HCC)}, pages 235--238. IEEE, 2023.

\bibitem{lill2024helpfulness}
A.~Lill, A.~N. Meyer, and T.~Fritz.
\newblock On the helpfulness of answering developer questions on discord with similar conversations and posts from the past.
\newblock In {\em Proceedings of the International Conference on Software Engineering}, pages Epub--ahead. ACM Digital library, 2024.

\bibitem{lima2023looking}
M.~Lima, I.~Steinmacher, D.~Ford, E.~Liu, G.~Vorreuter, T.~Conte, and B.~Gadelha.
\newblock Looking for related posts on github discussions.
\newblock {\em PeerJ Computer Science}, 9:e1567, 2023.

\bibitem{Lin2016}
B.~Lin, A.~Zagalsky, M.-A. Storey, and A.~Serebrenik.
\newblock Why developers are slacking off: Understanding how software teams use slack.
\newblock In {\em Proceedings of the 19th Conference on Computer Supported Cooperative Work and Social Computing Companion}, pages 333--336, 2016.

\bibitem{liu2018recommender}
C.~Liu, D.~Yang, X.~Zhang, H.~Hu, J.~Barson, and B.~Ray.
\newblock A recommender system for developer onboarding.
\newblock In {\em Proceedings of the 40th International Conference on Software Engineering: Companion Proceeedings}, pages 319--320, 2018.

\bibitem{liu2020bugsum}
H.~Liu, Y.~Yu, S.~Li, Y.~Guo, D.~Wang, and X.~Mao.
\newblock Bugsum: Deep context understanding for bug report summarization.
\newblock In {\em Proceedings of the 28th International Conference on Program Comprehension}, pages 94--105, 2020.

\bibitem{malheiros2012source}
Y.~Malheiros, A.~Moraes, C.~Trindade, and S.~Meira.
\newblock A source code recommender system to support newcomers.
\newblock In {\em Proceedings of the IEEE 36th annual computer software and applications conference}, pages 19--24. IEEE, 2012.

\bibitem{Mani2012}
S.~Mani, R.~Catherine, V.~S. Sinha, and A.~Dubey.
\newblock {AUSUM}: Approach for unsupervised bug report summarization.
\newblock In {\em Proceedings of the 20th International Symposium on the Foundations of Software Engineering}, pages 11:1--11:11, 2012.

\bibitem{McBurney2014}
P.~W. McBurney and C.~McMillan.
\newblock Automatic documentation generation via source code summarization of method context.
\newblock In {\em Proceedings of the 22nd International Conference on Program Comprehension}, pages 279--290, 2014.

\bibitem{Moreno2013}
L.~Moreno, J.~Aponte, G.~Sridhara, A.~Marcus, L.~Pollock, and K.~Vijay-Shanker.
\newblock Automatic generation of natural language summaries for {Java} classes.
\newblock In {\em Proceedings of the 21st International Conference on Program Comprehension}, pages 23--32, 2013.

\bibitem{Niu2013}
N.~Niu, A.~Mahmoud, Z.~Chen, and G.~Bradshaw.
\newblock Departures from optimality: Understanding human analyst's information foraging in assisted requirements tracing.
\newblock In {\em Proceedings of the International Conference on Software Engineering}, pages 572--581, 2013.

\bibitem{Parnin2011}
C.~Parnin and C.~Treude.
\newblock Measuring {API} documentation on the web.
\newblock In {\em Proceedings of the 2nd International Workshop on Web 2.0 for Software Engineering}, pages 25--30, 2011.

\bibitem{Pinto2016}
G.~Pinto, I.~Steinmacher, and M.~A. Gerosa.
\newblock More common than you think: An in-depth study of casual contributors.
\newblock In {\em Proceedings of the 23rd International Conference on Software Analysis, Evolution, and Reengineering}, pages 112--123, 2016.

\bibitem{Rastkar2014}
S.~Rastkar, G.~C. Murphy, and G.~Murray.
\newblock Automatic summarization of bug reports.
\newblock {\em IEEE Transactions on Software Engineering}, 40(4):366--380, 2014.

\bibitem{Rastogi2015}
A.~Rastogi, S.~Thummalapenta, T.~Zimmermann, N.~Nagappan, and J.~Czerwonka.
\newblock Ramp-up journey of new hires: Tug of war of aids and impediments.
\newblock In {\em Proceedings of the International Symposium on Empirical Software Engineering and Measurement}, pages 1--10, 2015.

\bibitem{rehman2020newcomer}
I.~Rehman, D.~Wang, R.~G. Kula, T.~Ishio, and K.~Matsumoto.
\newblock Newcomer candidate: Characterizing contributions of a novice developer to github.
\newblock In {\em Proceedings of the IEEE international conference on software maintenance and evolution (ICSME)}, pages 855--855. IEEE, 2020.

\bibitem{robillard2017demand}
M.~P. Robillard, A.~Marcus, C.~Treude, G.~Bavota, O.~Chaparro, N.~Ernst, M.~A. Gerosa, M.~Godfrey, M.~Lanza, M.~Linares-V{\'a}squez, et~al.
\newblock On-demand developer documentation.
\newblock In {\em Proceedings of the International Conference on Software Maintenance and Evolution}, pages 479--483. IEEE, 2017.

\bibitem{serrano2021find}
L.~P. Serrano~Alves, I.~S. Wiese, A.~P. Chaves, and I.~Steinmacher.
\newblock How to find my task? chatbot to assist newcomers in choosing tasks in oss projects.
\newblock In {\em Proceedings of the International Workshop on Chatbot Research and Design}, pages 90--107. Springer, 2021.

\bibitem{Sillito2006}
J.~Sillito, G.~C. Murphy, and K.~De~Volder.
\newblock Questions programmers ask during software evolution tasks.
\newblock In {\em Proceedings of the 14th International Symposium on Foundations of Software Engineering}, pages 23--34, 2006.

\bibitem{Steinmacher2015}
I.~Steinmacher, T.~Conte, M.~A. Gerosa, and D.~Redmiles.
\newblock Social barriers faced by newcomers placing their first contribution in open source software projects.
\newblock In {\em Proceedings of the 18th Conference on Computer Supported Cooperative Work \& Social Computing}, pages 1379--1392, 2015.

\bibitem{Steinmacher2016}
I.~Steinmacher, T.~Conte, C.~Treude, and M.~A. Gerosa.
\newblock Overcoming open source project entry barriers with a portal for newcomers.
\newblock In {\em Proceedings of the 38th International Conference on Software Engineering}, pages 273--284, 2016.

\bibitem{steinmacher2018let}
I.~Steinmacher, C.~Treude, and M.~A. Gerosa.
\newblock Let me in: Guidelines for the successful onboarding of newcomers to open source projects.
\newblock {\em IEEE Software}, 36(4):41--49, 2018.

\bibitem{Steinmacher2014}
I.~Steinmacher, I.~S. Wiese, T.~Conte, M.~A. Gerosa, and D.~Redmiles.
\newblock The hard life of open source software project newcomers.
\newblock In {\em Proceedings of the 7th International Workshop on Cooperative and Human Aspects of Software Engineering}, pages 72--78, 2014.

\bibitem{Storey2010}
M.-A. Storey, C.~Treude, A.~van Deursen, and L.-T. Cheng.
\newblock The impact of social media on software engineering practices and tools.
\newblock In {\em Proceedings of the FSE/SDP Workshop on Future of Software Engineering Research}, pages 359--364, 2010.

\bibitem{Treude2011b}
C.~Treude, O.~Barzilay, and M.-A. Storey.
\newblock How do programmers ask and answer questions on the web? ({NIER} track).
\newblock In {\em Proceedings of the 33rd International Conference on Software Engineering}, pages 804--807, 2011.

\bibitem{Treude2012}
C.~Treude, F.~{Figueira Filho}, B.~Cleary, and M.-A. Storey.
\newblock Programming in a socially networked world: the evolution of the social programmer.
\newblock In {\em Proceedings of the CSCW Workshop on the Future of Collaborative Software Development}, pages 1--3, 2012.

\bibitem{Treude2015}
C.~Treude, F.~Figueira~Filho, and U.~Kulesza.
\newblock Summarizing and measuring development activity.
\newblock In {\em Proceedings of the 10th Joint Meeting on Foundations of Software Engineering}, pages 625--636, 2015.

\bibitem{Krogh2003}
G.~Von~Krogh, S.~Spaeth, and K.~R. Lakhani.
\newblock Community, joining, and specialization in open source software innovation: a case study.
\newblock {\em Research Policy}, 32(7):1217--1241, 2003.

\bibitem{xiao2024devgpt}
T.~Xiao, C.~Treude, H.~Hata, and K.~Matsumoto.
\newblock Devgpt: Studying developer-chatgpt conversations.
\newblock In {\em Proceedings of the International Conference on Mining Software Repositories}, 2024.

\bibitem{xiao2023personalized}
W.~Xiao, J.~Li, H.~He, R.~Qiu, and M.~Zhou.
\newblock Personalized first issue recommender for newcomers in open source projects.
\newblock In {\em Proceedings of the 38th IEEE/ACM International Conference on Automated Software Engineering (ASE)}, pages 800--812. IEEE, 2023.

\bibitem{zhang2023ga}
M.~Zhang, G.~Zhou, W.~Yu, N.~Huang, and W.~Liu.
\newblock Ga-scs: Graph-augmented source code summarization.
\newblock {\em ACM Transactions on Asian and Low-Resource Language Information Processing}, 22(2):1--19, 2023.

\end{thebibliography}
\end{document}